\documentclass[aps,pra,twocolumn,a4paper,superscriptaddress,floatfix,10pt,longbibliography,nopacs]{revtex4-2}
\usepackage{graphicx}
\usepackage{dcolumn}
\usepackage{bm}
\usepackage{color}
\usepackage{amsmath}
\usepackage{amssymb}
\usepackage{hyperref}

\begin{document}

\title{Itinerant ferromagnetism in the repulsive Hubbard chain with anisotropic odd-wave attraction}
\author{M. Singh}
\affiliation{Universit\' e de Paris, Laboratoire Mat\' eriaux et Ph\' enom\`enes Quantiques, CNRS, F-75013, Paris, France}
\author{S. Pilati}
\affiliation{School of Science and Technology, Physics Division, Universit\`a di Camerino, 62032 Camerino (MC), Italy}
\author{G. Orso}
\email{giuliano.orso@univ-paris-diderot.fr}
\affiliation{Universit\' e de Paris, Laboratoire Mat\' eriaux et Ph\' enom\`enes Quantiques, CNRS, F-75013, Paris, France}

%
%

\begin{abstract}
The ground-state properties of the Hubbard chain with on-site repulsion and anisotropic nearest-neighbor attraction 
 are investigated by means of density matrix renormalization group calculations. The non-local attraction acts between fermions of one spin component only, mimicking the effect of p-wave Feshbach resonances in cold-atom systems.
%
We analyze the onset of itinerant ferromagnetism, pinpointing the critical attraction strength where partially and fully ferromagnetic states occur.
In the cold-atom setup, where the two (pseudo) spin populations are separately conserved, ferromagnetism occurs with the nucleation of a fully imbalanced band-insulating domain hosting the attractive component only. The size of this domain grows with the attraction strength, therefore increasing the (opposite) imbalance of the other domain, until the two spin components are fully separated.
In the presence of a harmonic trap, the ferromagnetic state hosts a partially imbalanced domain in the center with an excess of the attractive component and filling lower than one. This central region is surrounded by fully imbalanced domains, located in the trap tails, hosting only fermions belonging to the other component.
\end{abstract}

\maketitle

\section{Introduction}
In various metals, such as nickel, cobalt, and iron, itinerant electrons display ferromagnetic behavior.
The first theory introduced to characterize this phenomenon, which is usually referred to as itinerant ferromagnetism, is Stoner's 1933 continuous-space  Hamiltonian~\cite{stoner}.
Also the celebrated Hubbard model, which describes electrons hopping between the sites of a discrete lattice, was originally introduced to characterize itinerant ferromagnetism~\cite{hubbard}.
However, it is still unclear if and when the conventional Hubbard model -- i.e., the one including only nearest-neighbor hopping and on-site repulsion -- has a ferromagnetic ground-state~\cite{RUDIN1985273,hirsch1985two}, beyond the infinite-interaction limit~\cite{nagaoka1966ferromagnetism,carleo}.
The important role of additional hopping and interaction terms has been stressed in the condensed-matter literature (see, e.g., Ref.~\cite{amadon}).

In recent years, cold-atom experiments have emerged as the ideal platform to investigate quantum magnetism in strongly correlated systems.
In particular, deep optical lattices have allowed the implementation of Hubbard-type Hamiltonians~\cite{jaksch2005cold}.
Antiferromagnetism has been unambiguously observed in deep optical lattices close to half-filling~\cite{mazurenko2017cold,boll2016spin,cheuk2016observation,drewes2017antiferromagnetic}.
Following early attempts~\cite{jo2009itinerant,ketterle2012,ketterle2012prl}, signatures of itinerant ferromagnetism have been observed too~\cite{valtolina2017exploring,PhysRevLett.121.253602}. These latter experiments, performed  in a setup without an optical lattice, addressed the metastable upper branch of a resonantly interacting Fermi gas. The results were consistent with continuous-space quantum Monte Carlo simulations~\cite{PhysRevLett.103.207201,pilati2010,Chang51}.
Several procedures have been proposed to shift the onset of itinerant ferromagnetism to weaker interactions. This would allow experimentalists to avoid the three-body collisions that plague the strongly-interacting regime and prevent the creation of a more stable ferromagnetic state.
The list of proposed procedures includes: tuning the interaction strength via narrow Feshbach resonances~\cite{kohstall2012metastability,massignan2011repulsive}, adding shallow optical lattices~\cite{dft,pilati2014}, optical-flux lattices~\cite{PhysRevLett.109.265301}, flat-band optical lattices~\cite{PhysRevA.82.053618} or correlated disorder~\cite{pilati2016ferromagnetism}, using atomic species with different masses~\cite{PhysRevLett.110.165302,fratini2014zero}, and trapping atoms in confined low-dimensional geometries~\cite{Matveeva,volosniev2014strongly,PhysRevA.87.060502,PhysRevLett.111.045302,Koutentakis_2020}.
More recently, it has been proposed to favor itinerant ferromagnetism by means of an attractive intra-species interaction, tuned via a p-wave Feshbach resonance~\cite{jiang2016itinerant,yang2016engineering,Kurlov:PRA2019}. 
This mechanism has been studied only in one-dimensional continuous-space models and its generalization to strongly correlated lattice systems, including atoms confined in deep optical lattices, remains an open problem. Furthermore, it is not clear how phase-separation would occur 
for bulk systems in the standard cold-atom setup, where the (pseudo) spin populations are separately conserved. 

In this Article, we investigate the ground-state properties of a repulsive Hubbard chain augmented with an anisotropic nearest-neighbor intra-species attraction. Our calculations are based on the numerically-exact density matrix renormalization group (DMRG) technique~\cite{schollwock2005density}.
Due to the splitting between the Feshbach resonances in the triplet-states with different spin-projections, p-wave resonant interactions break spin-rotation symmetry~\cite{PhysRevLett.90.053201,PhysRevA.69.042712}. To describe this scenario, our model includes attractive interactions between fermions of one spin  component only.
Due to the presence of SU(2) symmetry-breaking interactions, the Lieb-Mattis theorem~\cite{lieb2002theory}, stating that the ground state of the one dimensional Hubbard model is a singlet, does not apply. Therefore, a ferromagnetic ground-state is in principle possible even in one-dimension~\cite{yang2004ferromagnetic}.
To determine if and when ferromagnetism occurs, we determine the ground-state energy as a function of the attraction strength and of the spin-population imbalance. The spin-resolved density profiles as well as the double occupancy are computed.
The critical point where ferromagnetism occurs is pinpointed, considering both the typical condensed matter setup where spin-rotation mechanisms are present, and also the standard cold-atom setup where the two (pseudo) spin populations are individually conserved, thus fixing the global spin polarization.
In the latter case, ferromagnetism manifests as a phase separation. At the critical point, a band-insulating domain appears, hosting the attractive species only. The size of this domain grows with the attraction strength, therefore increasing the (opposite) population imbalance in the other domain, until the two spin components are fully separated.
Finally, we show that the presence of an additional harmonic trap leads to a different scenario for the formation of the ferromagnetic state. In particular, the band-insulating  domain of the attractive species emerges in the middle of the trap only after 
fully polarized domains of the opposite spin component have formed in the trap tails.

The rest of the Article is organized as follows:  the Hamiltonian we study is described in Section~\ref{modelmethods}, together with some details on the DMRG technique. The results for the ground-state energy, the spin-resolved density profiles, the double occupancy, as well as the analysis of the onset of ferromagnetic behavior, are presented in Section~\ref{results}. Section~\ref{conclusions} reports a summary of our main findings.

\section{Model and computational details}
\label{modelmethods}
We consider a one dimensional spin-1/2 Fermi gas described by the following generalized Hubbard Hamiltonian:
\begin{eqnarray}
H&=&-t \sum_{i\sigma}(c_{i\sigma}^{\dagger} c_{i+1 \sigma}+h.c. )+ U \sum_{i}  n_{\uparrow i}n_{\downarrow i} \nonumber \\
&+&  \sum_{i\sigma\sigma^\prime}  V^{\sigma\sigma^\prime} n_{\sigma i}n_{\sigma^\prime i+1},
\label{eq:zero}
\end{eqnarray}
where $c_{i\sigma}^{\dagger} (c_{i\sigma})$ and $n_{\sigma i}$ are the creation (annihilation) and number operators of fermions
with spin projection ${\sigma=\uparrow,\downarrow}$, $t$ is the hopping rate between 
nearest neighboring sites, and $U$ is the strength of the on-site repulsive interaction between fermions with opposite spins. The model~\eqref{eq:zero} includes additional spin-dependent nearest-neighbor interactions of strength $V^{\sigma\sigma^\prime}$, mimicking the effect of odd-wave (specifically p-wave) anisotropic interactions. Since the latter are only relevant for fermion pairs with a given (total) spin projection, we consider 
%
 %
 %
$V^{\sigma\sigma^\prime}=V$ for $\sigma=\sigma^\prime=\uparrow$ and  $V^{\sigma\sigma^\prime}=0$ otherwise. 
Importantly, we assume that the nearest neighbor interactions are attractive in nature, corresponding to $V<0$, so that they might favor ferromagnetism. 
%
In this Article, the Hubbard chain is considered with open boundary conditions, corresponding to the configuration of a flat box with hard wall boundaries. This type of configuration can be created in cold-atoms experiments using almost uniform traps, as in Refs.~\cite{gaunt2013bose,mukherjee2017homogeneous,hueck2018two}. Below, we will also consider the addition of a harmonic confinement, which describes the effect of more conventional magneto-optical traps.

The ground state properties of the Hamiltonian~\eqref{eq:zero} are investigated using the DMRG method, expressed in terms of matrix
product states~\cite{schollwoeck2011}. Specifically, we use the open-source  code of
the ALPS library~\cite{dolfi2014}. 
To ensure proper convergence of the various observables, we allow for bond dimensions up to $4000$ and perform a large number of sweeps (between $60$ and $80$). 

\section{Results}
\label{results}

\begin{figure}
\includegraphics[width=\columnwidth]{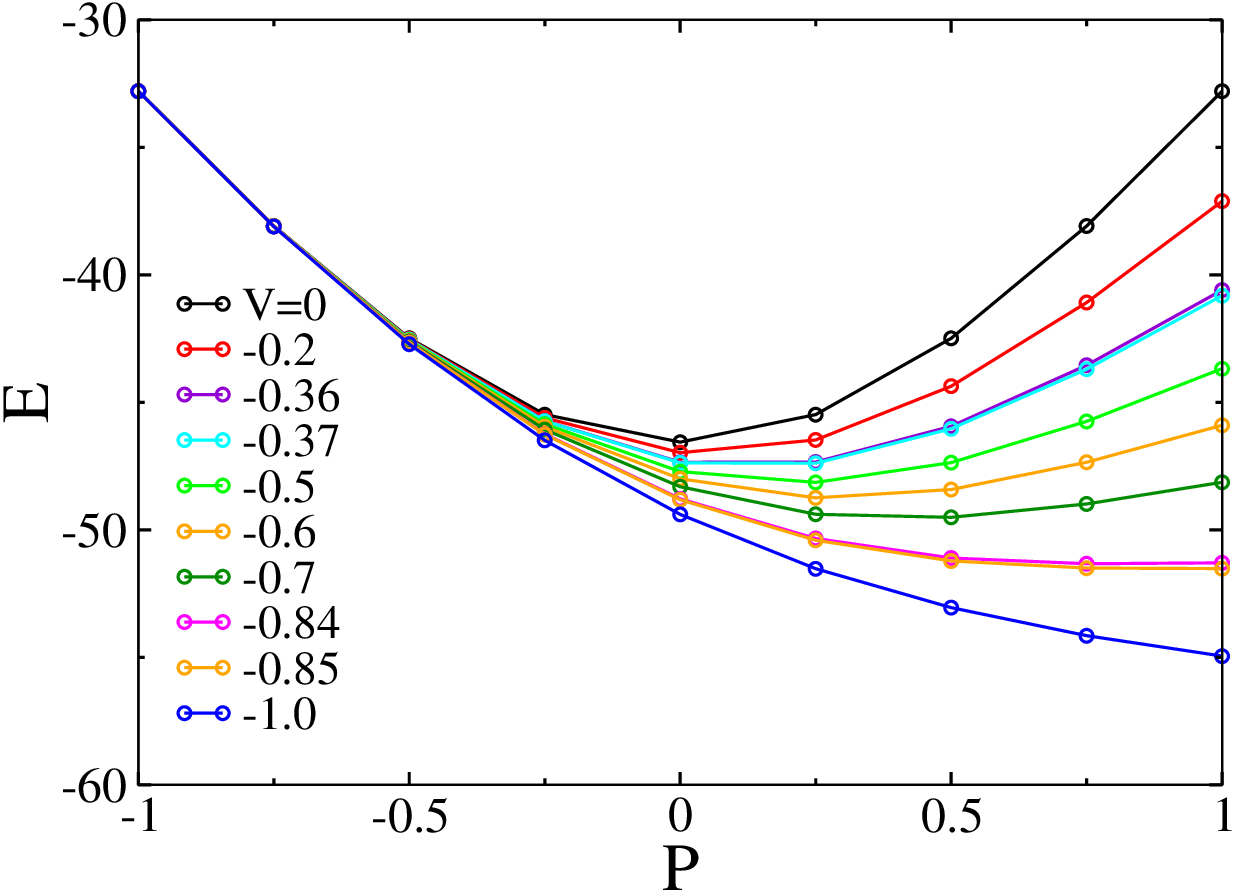}
\caption{(Color online) Ground-state energy $E$ as a function of the spin polarization $P=(N_\uparrow-N_\downarrow)/(N_\uparrow+N_\downarrow)$, plotted for different values of the interaction strength $V$, ranging from $V=0$ (upper curve) to $V=-1$ (bottom curve). 
The total number of fermions is fixed to $N=N_\uparrow+N_\downarrow=40$, the length of the chain is $L=60$. Here and in all other figures the on-site Hubbard-repulsion strength is $U=5$; furthermore, the connecting lines are a guide to the eye, unless otherwise specified.}
\label{fig:fig1}
\end{figure}

Hereafter, we fix the energy scale by setting $t=1$, while the strength of the on-site repulsion is fixed  to $U=5$, corresponding to the strongly interacting regime.

Our main aim is to shed light on the effect of the nearest neighbor attraction on onset and on the stability of itinerant  ferromagnetism.
%
In Fig.~\ref{fig:fig1} we show the ground-state energy as a function of the spin polarization $P=(N_\uparrow-N_\downarrow)/(N_\uparrow+N_\downarrow)$, for different values of the attraction strength $V$.
 These data are obtained by keeping the total number of particles constant at $N=N_\uparrow+N_\downarrow=40$. The length of the chain is  $L=60$, so that the total density of fermions is $N/L=2/3$. 
For finite $V$, the ground state
energy is no longer a symmetric function of the spin polarization.  Since the nearest neighbor attraction affects the spin-up component only,
its effect on the ground state energy is more sizable for positive spin polarizations, where it diminishes the energy.
For $V>-0.365$, the ground state energy has a minimum at $P=0$, indicating that the ground state is paramagnetic. 
For stronger attraction, however, the minimum shifts to positive values of the spin polarization. 
If spin-rotating processes were present, as in typical condensed matter systems, the ground state would turn (partially) ferromagnetic, with a spin polarization corresponding to the position of the energy minimum.
By further increasing the nearest neighbor attraction, so that 
$V\lesssim-0.85$, the minimum of the energy occurs at $P=1$, implying that the ground state is fully ferromagnetic.

\begin{figure}
\includegraphics[width=\columnwidth]{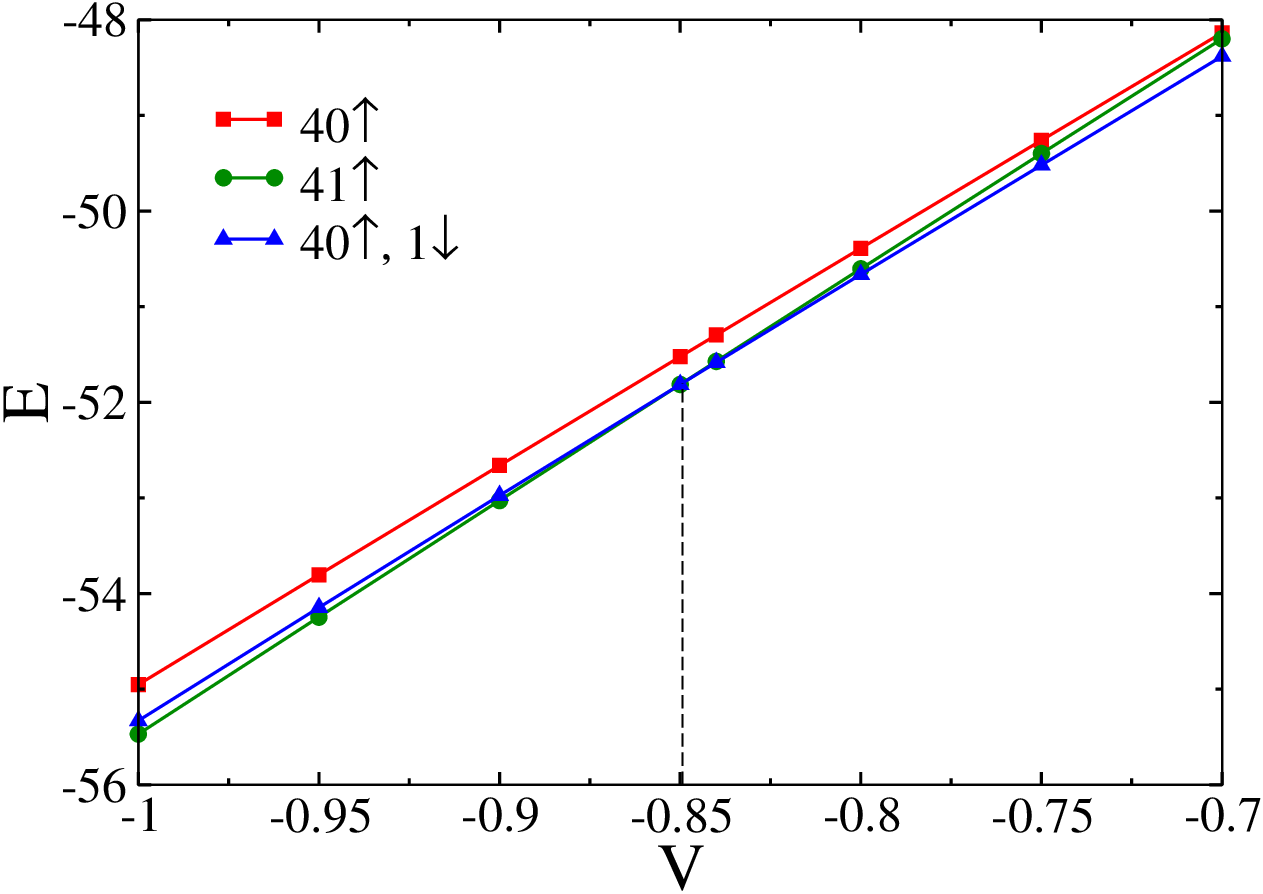}
\caption{(Color online)
Ground-state energy $E$ as a function of the nearest neighbor attraction strength $V$, for three different sets of population numbers. The red squares  correspond to a fully polarized gas of spin-up fermions, with $N_\uparrow=40$ and $N_\downarrow=0$. The other two datasets represent the energy of the same system upon addition of, respectively, a further spin-up fermion (green circles) or a 
spin-down fermion (blue triangles). The dashed vertical segment indicates the position of the crossing point, $V=-0.849(1)$. This corresponds to the critical value of the attraction strength for the onset of full ferrromagnetism.
The length of the chain is $L=60$. }
\label{fig:fig2}
\end{figure}

To precisely determine the critical attraction strength where the fully ferromagnetic phase occurs, we analyze the energy cost of adding a spin-up or a spin-down fermion to a gas of spin-up fermions only.
As a reference, in Fig.~\ref{fig:fig2} we show the ground-state energy of a fully polarized gas of $N_\uparrow=40$ fermions, plotted as a function of the interaction strength $V$ (see red squares). 
The green and the blue lines correspond to the ground-state energies with an additional spin-up or a spin-down fermion, respectively. The two curves cross  at $V=-0.849(1)$. For more negative $V$, adding a spin-up fermion reduces the energy by a larger amount than adding a spin-down fermion, implying that the ground state is fully ferromagnetic.
It is worth reminding that the change in energy due to the inclusion of an extra fermion with spin $\sigma$ represents the chemical potential 
$\mu_\sigma$ of the corresponding spin component. 

\begin{figure}
\includegraphics[width=\columnwidth]{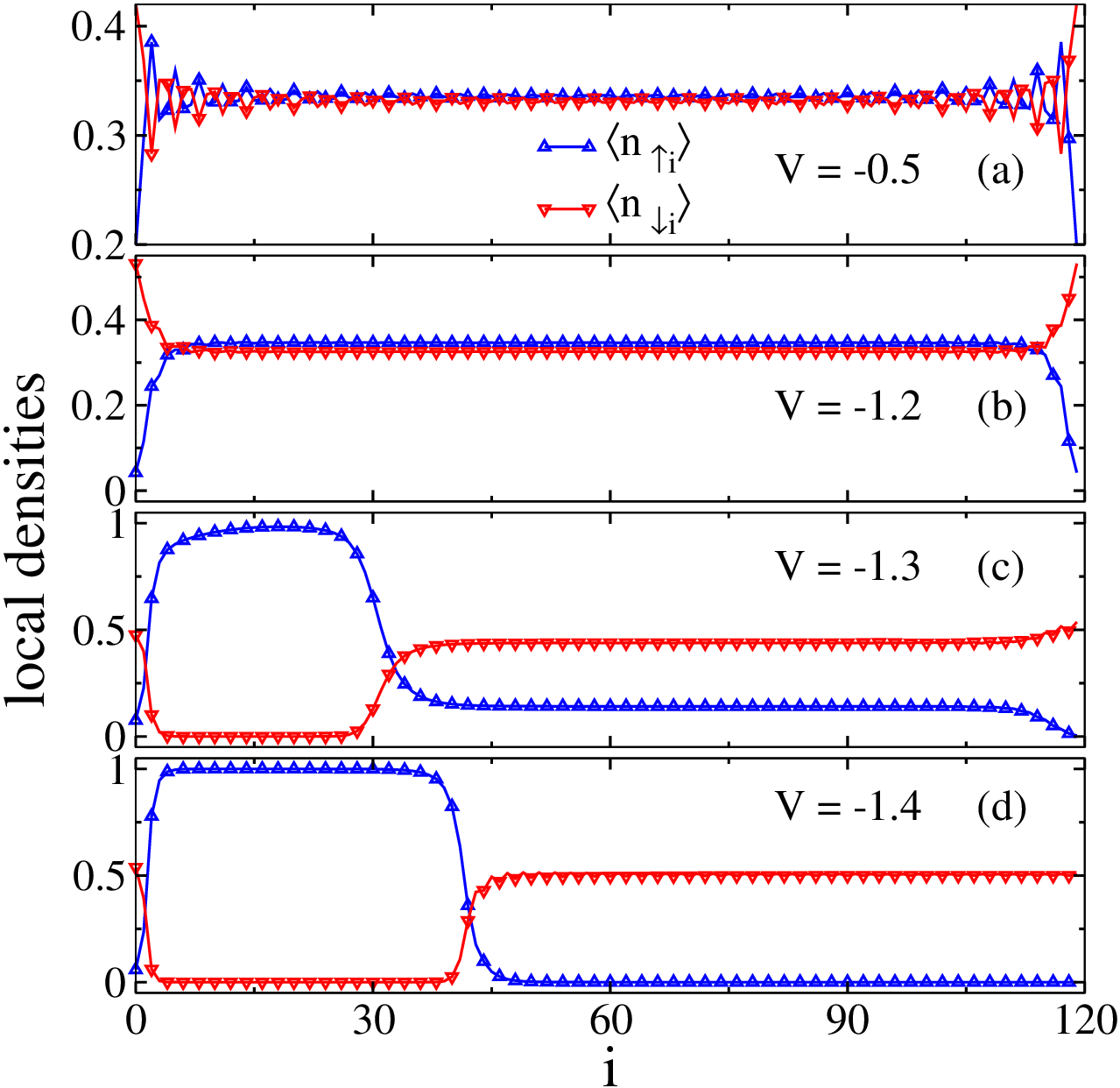}
\caption{(Color online) Density profiles of spin-up fermions (upward triangles) and spin-down fermions (downward triangles) in a chain with $L=120$ sites, labelled by the index $i$. 
To improve visibility, only every other point is shown.
The four panels (a)-(d) refer to different values of the nearest-neighbor attraction strength $V$ between spin-up fermions: $V=-0.5,-1.2,-1.3,-1.4 $ (from top to bottom). The spin populations are $N_\uparrow=N_\downarrow=40$. For large and negative $V$,  a growing fully-imbalanced domain hosting a band insulator of spin-up particles (i.e., the attractive component) coexists with a partially ferromagnetic domain with an excess of spin-down fermions.}
\label{fig:fig3}
\end{figure}

So far we have considered how itinerant ferromagnetism occurs assuming that the global spin polarization can vary to minimize the ground-state energy. Hereafter, we discuss the emergence of ferromagnetism under the assumption that the numbers of spin-up and spin-down fermions 
are fixed and coincide, $N_\uparrow=N_\downarrow=N/2$. Therefore, the global spin polarization is always zero. 
This is the common setup in cold-atom experiments, where the population of atoms in the two hyperfine states, which play the role of pseudo-spin components, are separately conserved.
In this setup, ferromagnetic phases correspond to 
phase-separated states hosting regions where the local spin densities are finite. 
To identify such states, we compute the density profiles of the two spin 
components.
In Fig. \ref{fig:fig3} we show the results for a chain of $L=120$ sites, filled with $N=80$ fermions. 
The four panels correspond to as many different values of the strength of the nearest neighbor attraction. 
For $V=-0.5$ the density profiles of the two spin components essentially coincide, indicating that the system is paramagnetic. The two profiles substantially differ only close to the walls, mainly due to the open boundary conditions and the on-site repulsion between the two spin components. One also observes small out-of-phase oscillations, analogous to Friedel oscillations, which are magnified near the system boundaries.
For $V=-1.2$ the Friedel-like oscillations disappear; the average densities of the two spin components  are only slightly different away from the boundaries, implying that the system is still paramagnetic.
For $V=-1.3$, the system is no longer homogeneous. A domain including only spin-up fermions with nearly unit local filling coexists with a partially imbalanced phase
hosting a majority of spin-down fermions. 
The domain with only spin-up fermions migrates towards one edge of the chain to further decrease the kinetic energy of the system.
Finally, for $V=-1.4$ the same domain has englobed all spin-up fermions, forming a band insulator occupying $N_\uparrow$ sites. The latter coexists with a fully polarized gas of spin-down fermions occupying the remaining sites, with average density $N_\downarrow/(L-N_\uparrow)=0.5$, as displayed in the bottom panel of Fig. \ref{fig:fig3}.

\begin{figure}
\includegraphics[width=\columnwidth]{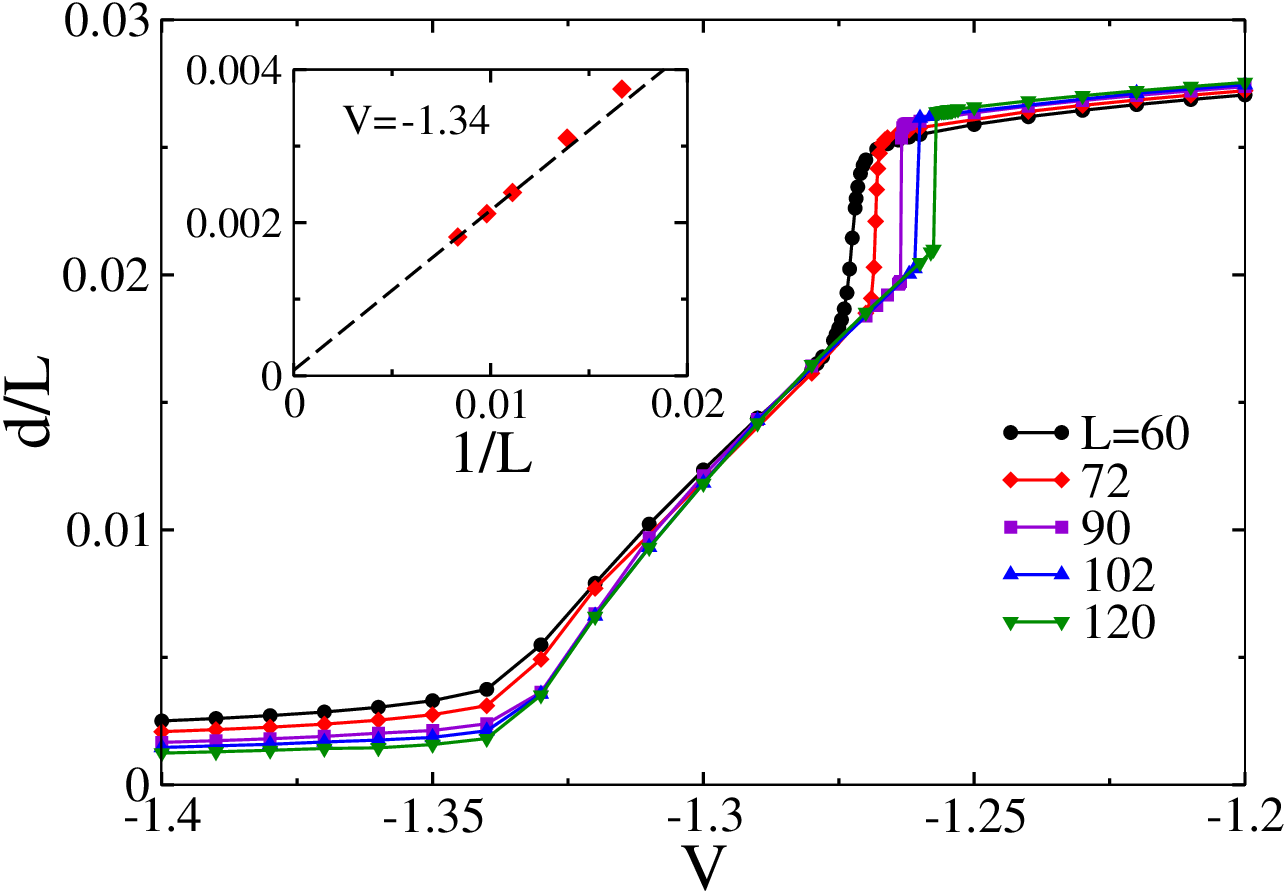}
\caption{(Color online) Double occupancy  $d$ as a function of the nearest neighbor attraction strength $V$. The different curves correspond to calculations for different system sizes.  For each value of $L$ we choose the spin populations so that the overall densities are kept constant to $N_\uparrow/L=N_\downarrow/L=1/3$, so that the global spin polarization is $P=0$. 
Notice the sudden variation around
$V=-1.25$, corresponding to the nucleation of the band-insulator domain hosting spin-up fermions only, leading to an essentially vertical drop of the double occupancy. The inset shows the  double occupancy as a function of $1/L$ for $V=-1.34$, showing that $d$ vanishes 
for infinite system sizes as the ground state becomes fully ferromagnetic. }
\label{fig:fig4}
\end{figure}

The emergence of the ferromagnetic phases  has a clear fingerprint also in the double occupancy $d=\sum_{i}  \langle n_{\uparrow i}n_{\downarrow i}\rangle/L $. This is plotted in  Fig.~\ref{fig:fig4} as a function of the attraction strength $V$, for different values of the system size. 
As expected,  $d$ is an increasing function of the nearest neighbor interaction strength $V$. 
For large and negative $V$, where the ground-state is fully ferromagnetic, the only contribution to the double occupancy comes from the overlapping tails of the two spin-polarized domains. Being a surface effect, this contribution vanishes in the thermodynamic limit.
 To verify this point, in the inset of Fig. \ref{fig:fig4} we plot the double occupancy  for $V=-1.34$ as a function of $1/L$. The dashed line corresponds to a linear fit of the data obtained by retaining only the three largest system sizes considered. The
 intercept is approximately zero, implying that the double occupancy vanishes in the thermodynamic limit.
 In contrast, we see from  Fig.~\ref{fig:fig4}  that for $V=-1.33$ the same quantity saturates to a finite value for large system sizes.
%
The value $V \cong -1.34$ can then be identified as the critical attraction strength for the onset of the fully ferromagnetic state under the constraint of zero global spin polarization. 
For intermediate values of $V$, where the fully polarized domain of spin-up fermions  coexists with the partially ferromagnetic phase, finite-size effects are almost negligible for the largest system sizes considered. At the interaction strength where such domain disappears, the double occupancy displays a sudden variation. 
For weaker nearest neighbor attractions, where the ground state is paramagnetic, the density profiles of the two spin components overlap over the entire chain, leading to significantly larger values of the double occupancy. In contrast with what observed for the fully ferromagnetic phase, here $d$ increases with system size.
This is due to the fact that the presence of the hard walls and the repulsive on-site interactions cause a depletion of the double occupancy. Being a surface effect, the depletion diminishes as $L$ increases. 

\begin{figure}
\includegraphics[width=\columnwidth]{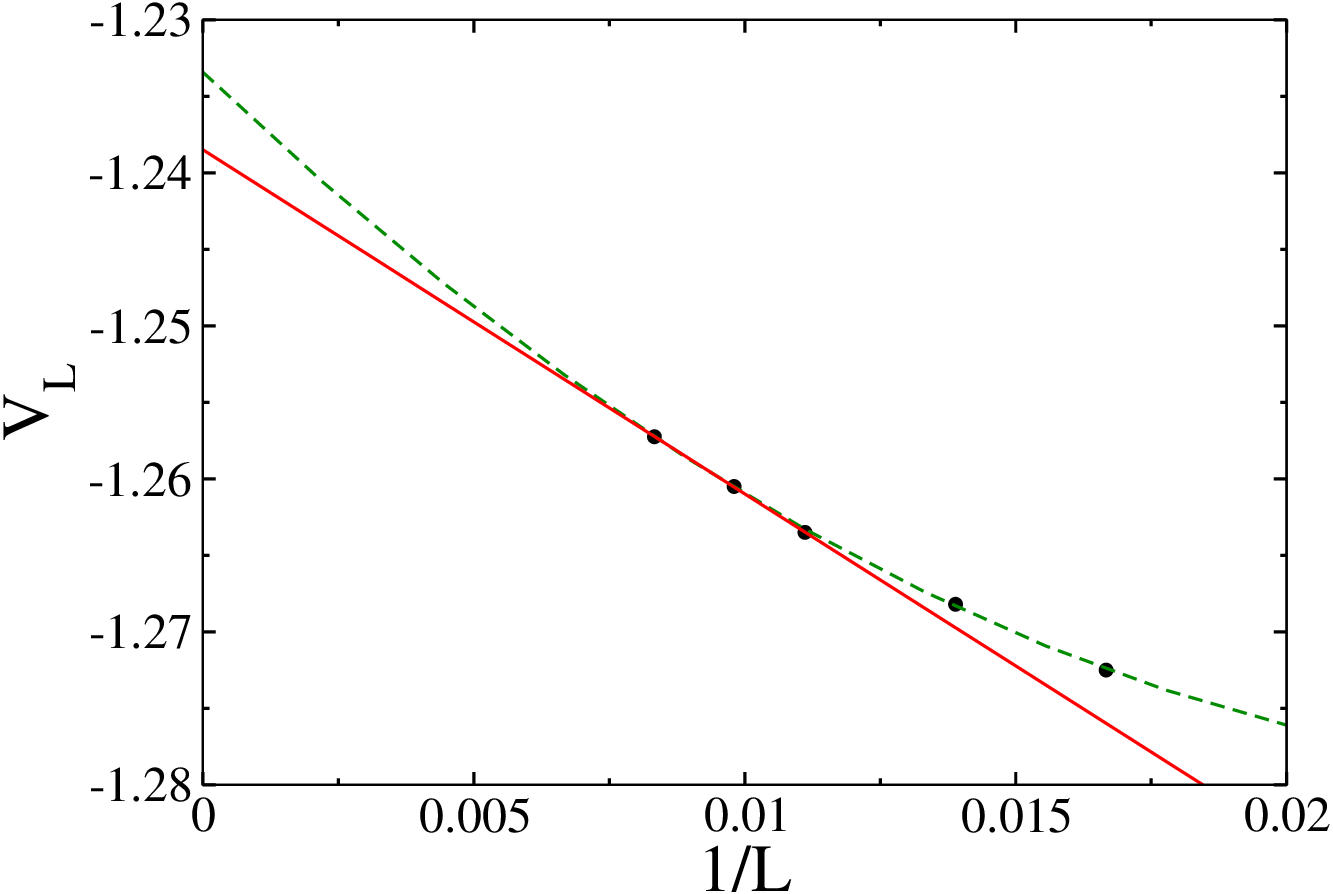}
\caption{Extrapolation to the infinite-size limit of the critical strength $V=V_c$ for the nucleation of the band-insulating domain, corresponding to the phase transition to a partially ferromagnetic state. For each system size $L$, we extract the value $V=V_L$ at which the double occupancy in Fig. \ref{fig:fig4} displays the sudden variation and plot it as a function of $1/L$. 
  The position of the critical point is inferred  by fitting the data with a low order polynomial $p(x)$, with  $x=1/L$, and setting $V_c=p(0)$. 
Specifically, a linear fitting function (continuous red line), obtained by retaining only the three largest system sizes, and a quadratic fitting function (dashed green curve) are shown, yielding 
$V_c=-1.2385(3)$ and $V_c=-1.233(2)$, respectively.}
\label{fig:fig5}
\end{figure}

 %
We also see from Fig.~\ref{fig:fig4} that the position of the sudden variation of the  double occupancy  shifts towards  weaker nearest neighbor attractions for larger system sizes, thus broadening the parameter region where the paramagnetic phase is unstable against phase separation. 
%
We can estimate the critical value of the interaction strength $V=V_c$ as the position of the sudden variation in the thermodynamic limit, $L\rightarrow +\infty$.  This value is determined by performing a finite-size scaling analysis.
For each system size $L$, we determine the value $V=V_L$ of the interaction strength at which the reduced double occupancy  in Fig. \ref{fig:fig4} exhibits an inflexion point in the critical region.
 For the largest system sizes, where the sudden variation is essentially a vertical jump, we simply identify $V_L$ with the position of the jump. 
 The obtained results are displayed  in  Fig. \ref{fig:fig5} as a function of the system inverse size $1/L$. The dashed continuous line corresponds to a linear fit of the data, $V_L=a+b/L$, obtained by retaining only the three largest system sizes. This gives a critical interaction strength  $V_c=a=-1.2385(3)$.  
 A closer look to the data reveals that  a quadratic fit $V_L=a+b/L+c/L^2$ to the entire data set is also plausible. The result is
 shown in Fig. \ref{fig:fig5} by the dot-dashed line, from which we get $V_c=a=-1.233(2)$. 
 The small difference between the linear and the quadratic extrapolations provides a confidence interval of the estimate of the critical point, which we finally quote as $V_c=-1.235(4)$.
The height of the vertical jump diminishes with the system size. However, our data do not allow us to ascertain if  it vanishes in the thermodynamic limit, or if it converges to a small but finite value. For this reason, we cannot unambiguously identify the order of the phase transition.
%
%
It is worth emphasizing that, in the cold-atom setup with fixed global spin polarization, the partially and the fully ferromagnetic states occur at significantly stronger attraction compared to the case where the system can vary the global spin polarization to minimize its ground-state energy.
Furthermore, one notices that the parameter region where the ground-state is partially ferromagnetic, namely $-1.34 \lesssim V \lesssim -1.235$, is quite narrow, indicating that the system rapidly transitions from the paramagnetic to the fully ferromagnetic phase.

\begin{figure}
\includegraphics[width=\columnwidth]{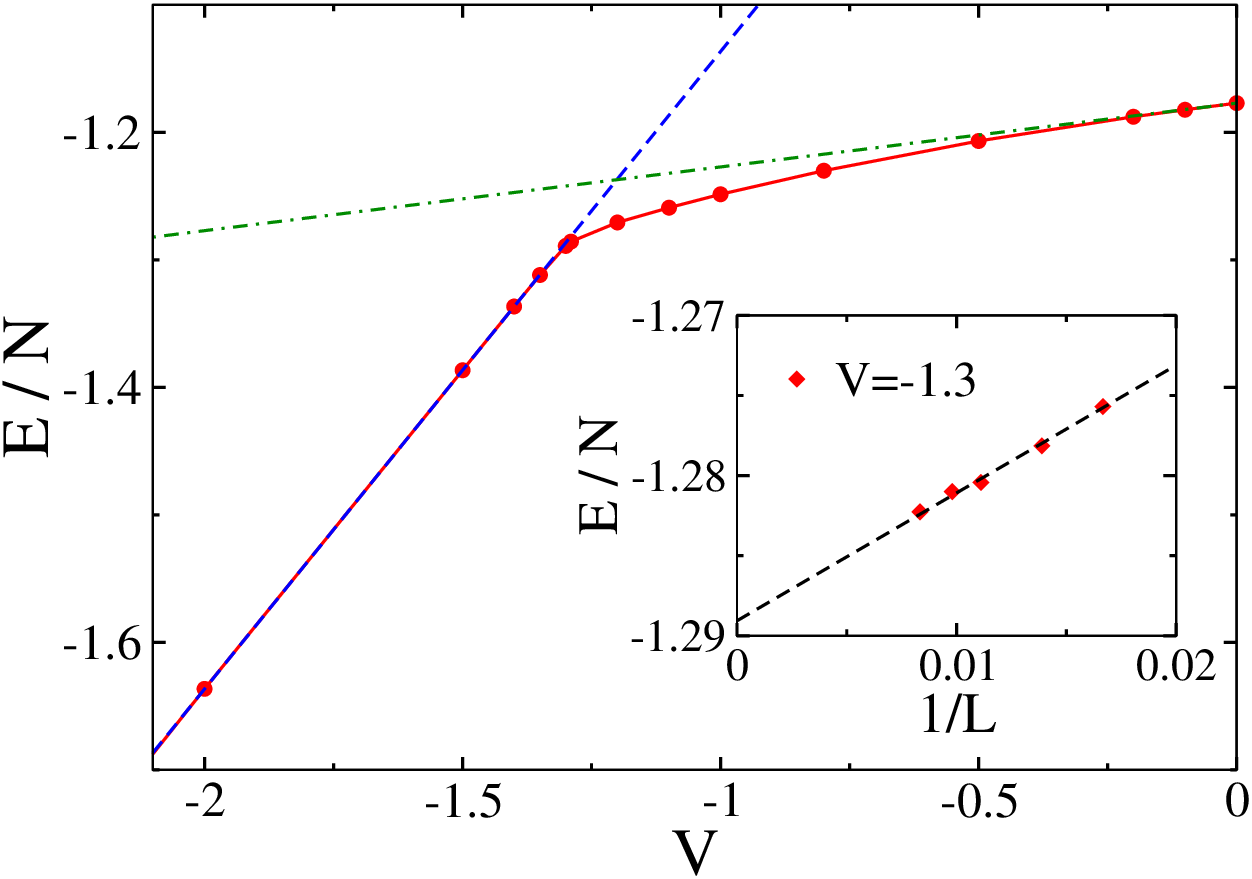}
\caption{Ground-state energy per particle $E/N$, extrapolated to the thermodynamic limit,  plotted as a function of the nearest neighbor attraction strength $V$. The asymptotic behavior
for large and negative $V$, see Eq.~\eqref{fullyferr}, is shown as a dashed line.  The dot-dashed line is the prediction from 
first order perturbation theory, holding for small and negative $V$. 
The inset shows the dependence of the ground-state energy per particle on the inverse system size (data symbol) for $V=-1.3$;
the spin densities are fixed at $N_\uparrow/L=N_\downarrow/L=1/3$.
The infinite-lattice result corresponds to the intercept of the straight line fitting the data (dashed curve).}
\label{fig:fig6}
\end{figure}

Next, we analyze the zero-temperature equation of state. We obtain the ground-state energy per particle $E/N$ 
in the thermodynamic limit via a linear extrapolation  as a  function of $1/L$.
 This is shown in the inset of  Fig. \ref{fig:fig6} for $V=-1.3$. By repeating the same procedure for the different values of the attraction 
 strength $V$, we obtain the curve shown in the main panel of Fig.~\ref{fig:fig6}.
%
%
The dashed and the dot-dashed straight lines correspond to the asymptotic behavior for strong and weak nearest neighbor attraction, respectively. 
In the regime of large negative $V$,
the band-insulating domain of the attractive spin-up fermions only coexists with the fully polarized ideal gas of spin-down fermions occupying the remaining $L-N_\uparrow$ lattice sites. Therefore, the ground-state energy per particle can be computed as
\begin{equation}\label{fullyferr}
E\simeq V \frac{(N_\uparrow-1)}{N}-\frac{2 (L-N_\uparrow)}{\pi N}\sin \frac{\pi N_\downarrow}{L-N_\uparrow}.
\end{equation}
 The asymptotic behavior~\eqref{fullyferr} is shown in Fig. \ref{fig:fig6} by the blue dashed line (we have neglected the $1/N$ term which vanishes in the thermodynamic limit). One notices that the prediction from Eq.~\eqref{fullyferr} is indeed very close to the DMRG result 
 for $V\lesssim -1.34$, where the ground state of the system is fully ferromagnetic. 
%
%
%
For small negative $V$, the effect of the nearest neighbor interactions in Eq.~\eqref{eq:zero} can be taken into account within first order perturbation theory, yielding $E\simeq E(V=0)+V\sum_{i}  \langle n_{\uparrow i}n_{\uparrow i+1}\rangle $, where the expectation value is computed for the Hubbard model, by assuming $V=0$ in Eq.~\eqref{eq:zero}. 
We observe that the perturbative behavior (green dot-dashed line) is only recovered for relatively small values of $V$.  
It is also worth noting that the two asymptotic lines cross at $V=-1.2$, which is not far from the critical point $V_c$ for the onset of 
the partially ferromagnetic phase through phase separation.

\begin{figure} 
\includegraphics[width=\columnwidth]{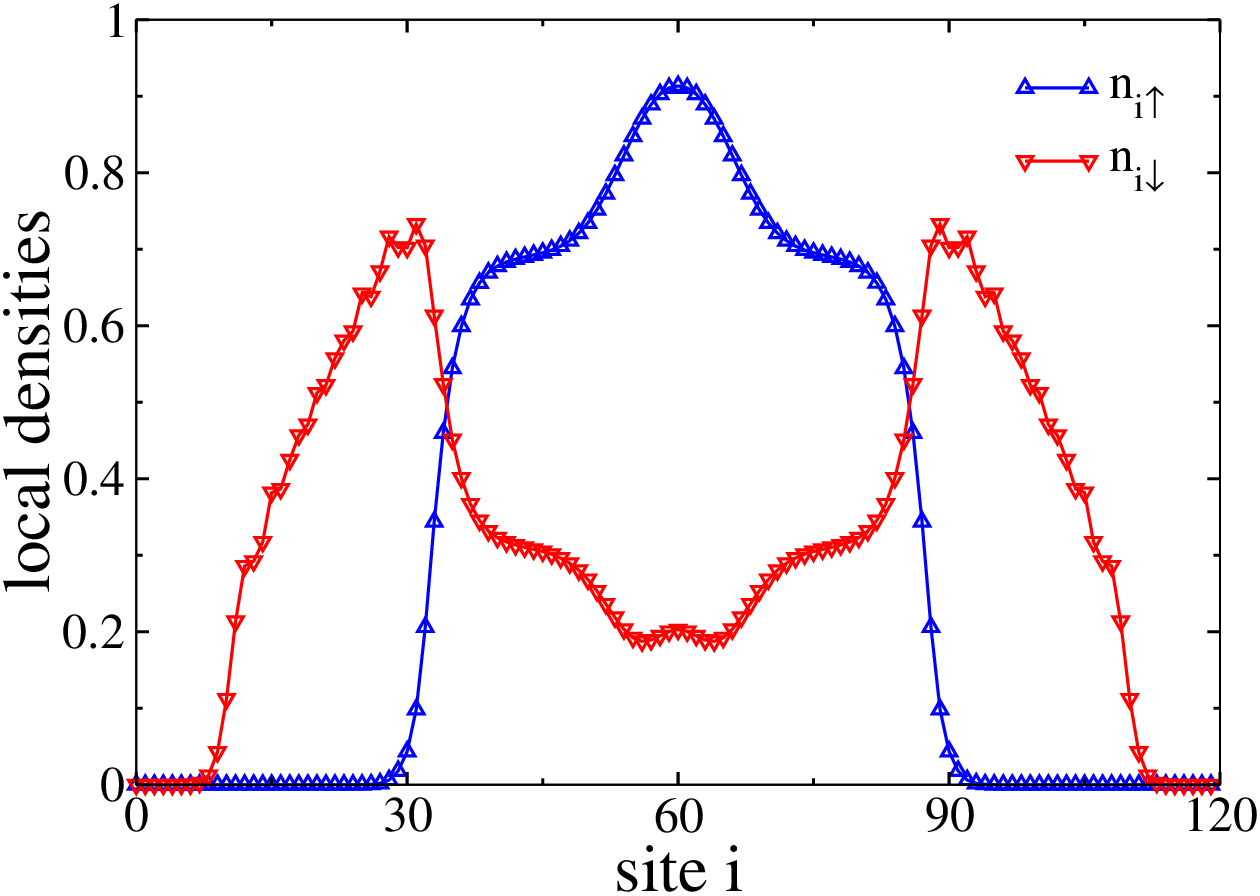}
\caption{(Color online) Density profiles of the two spin components in a harmonic trap of intensity $K=0.002$. The nearest-neighbor attraction strength is $V=-1.35$, which in the flat trap with open boundaries corresponds to a fully ferromagnetic ground state.  
The spin populations are $N_\uparrow=N_\downarrow=40$.}
\label{fig:fig7}
\end{figure}

Finally, we discuss how ferromagnetism forms when the Hamiltonian includes an additional longitudinal harmonic confinement. Specifically, we include the additional term:
\begin{equation}
H^\prime = \sum_{i} K \left (i-\frac{L}{2}\right )^2 (n_{i\uparrow}+n_{i\downarrow}), 
\end{equation}
where $K$ is a constant. This term is designed to describe the effect of the most common magneto-optical traps used to confine ultracold atoms. 
Fig.~\ref{fig:fig7} displays the spin-density profiles for $V=-1.35$. Notice that in the flat box trap without the harmonic confinement this attraction strength is sufficient to induce full phase separation of the two spin components. 
In the harmonic trap, the attractive spin-up fermions occupy mostly the trap center. This helps decreasing the interaction energy. Differently from the flat box case, where the spin-up domain is fully polarized, in the harmonic trap the (central) domain with the majority of spin-up fermions also hosts a lower density of spin-down fermions. The trap tails host fully polarized domains including spin-down fermions only, apart the small regions of interface with the central domain. For more negative $V$, say $V=-1.4$, a fully polarized 
spin-up domain emerges in the center of the trap, characterized by a flat density profile with unit filling.

\section{Conclusions}
\label{conclusions}
We have investigated the ferromagnetic properties of a Hubbard chain with on-site repulsion and anisotropic nearest-neighbor attraction between the spin-up fermions. 
The DMRG algorithm allowed us to compute  global properties such as the ground-state energy and the double occupancy, as well as local properties such as the spin-resolved density profiles. 
From energy calculations as a function of the spin-population imbalance we extracted the critical attraction strength for the transition to partially ferromagnetic and to fully ferromagnetic phases. By inspecting the density profiles and the double occupancy we determined how ferromagnetism occurs in the standard cold-atom setup where the (pseudo) spin populations are individually conserved. In this case, a significantly stronger attraction is needed to induce the separation of domains with non-zero local spin-population imbalance. Specifically, in the uniform system with open boundary conditions ferromagnetism occurs with the nucleation of a fully spin-polarized domain hosting spin-up fermions only. The size of this domain grows with the attraction strength until all spin-up fermions have been absorbed, meaning that the two spin components are fully separated.
The inclusion of an harmonic confinement substantially modifies the scenario. In this case, a partially imbalanced domain with an excess of spin-up fermions is located in the trap center, while the trap edges host domains with spin-down fermions only. In contrast, the band-insulating domain 
of spin-up fermions only is recovered for stronger nearest neighbor attractions.

We have presented an application of the DMRG algorithm to a novel Hamiltonian relevant to describe cold-atom systems. The computation of local properties allowed us to investigate the phase separation of different spin components and the coexistence of domains with different local spin-population imbalances. 
In particular, we have shown that the double occupancy, which is experimentally accessible with cold-atom systems, is a key quantity to investigate phase separation and the formation of different ferromagnetic phases.
Our findings can serve as a guide for future cold-atom experiments focussing on itinerant ferromagnetism in one-dimensional optical lattices with p-wave resonant interactions, and they complement previous studies on anti-ferromagnetic correlations in optical lattices~\cite{boll2016spin,hilker2017revealing,pilati2017one,salomon2019direct}. 

\section*{ACKNOWLEDGEMENTS} 
We acknowledge  fruitful discussions with A. Biella.
This work is supported by the ANR project SPIFBOX and by the SIRTEQ DIM program of the Region Ile-de-France through the project 1DFG. M. S. would like to acknowledge funding from MULTIPLY fellowships under the Marie Sk\l{}odowska-Curie COFUND Action 
(grant agreement No. 713694).
S. P. acknowledges financial support from the FAR2018 project titled ``Supervised machine learning for quantum
matter and computational docking'' of the University of Camerino and from the Italian MIUR under the project PRIN2017 CEnTraL 20172H2SC4.

\bibliography{Ref}

\end{document}